\documentclass[prb,twocolumn,showpacs,superscriptaddress,preprintnumbers,amsmath,amssymb,floatfix]{revtex4}

\usepackage{dcolumn}
\usepackage{graphicx}
\usepackage{bm} 

\newcommand{\eM}     {$\epsilon$-machine}
\newcommand{\eMs}    {$\epsilon$-machines}

\newcommand{\Range} {{$r$}}

\newcommand{\eMSR}    {{{$\epsilon$}MSR}}

\newcommand{\ACA}    {{ACA 232}}

\begin{document}

\title{From Finite to Infinite Range Order via Annealing:\\
The Causal Architecture of Deformation Faulting in\\
Annealed Close-Packed Crystals
}

\author{D. P. Varn}
\affiliation{Santa Fe Institute, 1399 Hyde Park Road, Santa Fe,
New Mexico 87501}

\author{J. P. Crutchfield}
\affiliation{Santa Fe Institute, 1399 Hyde Park Road, Santa Fe,
New Mexico 87501}

\date{\today}
\begin{abstract}
We analyze solid-state phase transformations that occur in zinc-sulfide
crystals during annealing using a random deformation-faulting mechanism with
a very simple interaction between adjacent close-packed double layers. We show that, through
annealing, infinite-range structures emerge from initially short-range crystal
order. That is, widely separated layers carry structurally significant
information and so layer stacking cannot be completely described by any
finite-range Markov process. We compare our results to two experimental
diffraction spectra, finding excellent agreement.
\end{abstract}

\pacs{
  61.72.Dd,   
  61.10.Nz,   
  61.43.-j,   
  81.30.Hd    
  \\
\begin{center}
  }

\maketitle

There has been considerable interest in understanding planar disorder in 
crystals for some time.~\cite{hendricks42,jagodzinski49a,pandey80a,pandey80b,pandey80c,berliner86,gosk01} 
There are several reasons for this emphasis.  
Since planar defects correspond to a shift of an entire layer of atoms that
nonetheless preserves the crystallinity within the layer, the disorder 
is confined along the stacking direction and results in a structure that  
may be treated as one-dimensional, making theoretical analysis tractable.~\cite{varn01a}   
Additionally, different stacking configurations can affect the physical 
properties of the material. For example, the band gap in SiC changes as the
stacking structure is changed~\cite{sebastian94} and an anomalous photovoltaic effect
has been observed in ZnS crystals that have disordered stacking sequences.~\cite{ellis58} 

While many crystals exhibit planar defects, they are especially common in a
class of materials known as {\em polytypes}.~\cite{sebastian94} Polytypism is
the phenomenon where a crystal is built up from the stacking of identical
two-dimensional layers, called {\em modular layers}
(MLs).~\cite{varn01a,varn02,varn03a,varn03b} Typically there is a small, finite set of possible  
ways a ML may be stacked upon another and, since these different stacking
orientations often preserve the atomic coordination number for nearest and
next-nearest neighbors, the energy difference between different stackings can 
be quite small. It is perhaps not surprising then that polytypic materials have many different
crystalline structures. There are, for instance, $185$ known crystalline
stacking structures in ZnS, some with unit cells extending over $100$
MLs.~\cite{sebastian94,trigunyat91} Other highly polytypic materials include
SiC, CdI$_2$, and AgI with about $150$, $200$, and $50$ known crystalline
structures each, respectively.~\cite{sebastian94}  

Understanding the variety and origin of spatial organization in crystalline
ZnS polytypes on length scales clearly in excess of the calculated inter-ML
interactions---$\sim$ 1 ML for ZnS~\cite{engel90b} and $\sim$ 3 MLs for
SiC~\cite{cheng87}---has been a puzzle for some time and numerous theories
have been proposed.~\cite{price84,yeomens88,frank51b,trigunyat91} Recently,
we extended this by demonstrating that {\em disordered} ZnS polytypes also
possess a long-range spatial organization in excess of the calculated
inter-ML interaction.~\cite{varn02,varn03b}  

In this paper we simulate transformations between ordered and disordered
polytypes, in particular, proposing a simple model to analyze the solid-state
transformation of annealed ZnS crystals from a 2H to a 3C structure and
finding a novel structural description---in the form of an
\eM~\cite{crutchfield89, Crut98d}---of the resulting disordered twinned
3C crystal. Our model is decidedly simpler than previous ones, designed
with the goal of determining the minimal complication necessary to produce
the experimentally observed long-range spatial order. For example, from
the \eM\ we show that infinite-range spatial memories arise even though the
interaction length between MLs is restricted to nearest neighbors. 
 
Disordered crystals are often formed by stressing a perfect crystal thermally,
mechanically, or through irradiation so that stacking faults are introduced.  
Various models have been proposed to explain these transformations. Typically,
the faulting process is assumed to proceed slowly, so that a ML is randomly
chosen and a fault is introduced if the local stacking configuration meets
some criteria, usually derived from a (theoretically or empirically determined)
Hamiltonian that describes the energetics of the inter-ML interactions. Further,
a particular faulting mechanism is assumed. 

Many authors have previously performed simulation studies on transformations in 
polytypes to understand the structure of faulted crystals.  
For example, by assuming interactions between MLs of up to a distance of three,
Kabra and Pandey~\cite{kabra88} were able to deduce that layer-displacement
faulting was the primary defect found in the 2H $\rightarrow$ 6H~\cite{note1}
transformation in SiC. They also discovered that this faulting mechanism
can result in long-range correlations between MLs without short-range order.
Engel~\cite{engel90a} applied a similar model to treat the 2H $\rightarrow$ 3C
transformation in ZnS via deformation faulting, with the assumption of
next-nearest neighbor interactions. These authors were able find stacking
configurations that gave x-ray diffractograms qualitatively similar to
experimental ones. 
Shrestha {\it et al.}~\cite{shrestha96a} treated the 2H $\rightarrow$ 3C 
martensitic transformation
in hypothetical close-packed structures by introducing the infinitely
strong repulsion model. They assumed that the presence of a fault inhibited
the introduction of another fault on adjacent MLs. That is, that there was some 
coordination between faults, called {\em nonrandom faulting}.~\cite{sebastian87a,sebastian87b} 
Gosk~\cite{gosk00,gosk01} developed a model to describe disordered 2H crystals 
that used a probability function for faulting on the next ML as one scans the
crystal that depended on the distance from the last observed fault.
By tuning various model parameters one could introduce nonrandom faulting into
the stacking sequence.  

ANNNI models~\cite{yeomens88} have also been used to explain polytypism since  
they are known to give long-range spatial organization with only short-range
interactions (up to next-nearest neighbor). However, ANNNI models require
fine tuning the coupling parameters in an interaction Hamiltonian and, thus,
may be of limited applicability to polytypes.  

ZnS has two stable phases: the hexagonal close-packed (or
2H)~\cite{varn03a,varn03b} for temperatures above 1024 C and the cubic
closed-packed (or 3C) for temperatures below 1024 C.~\cite{sebastian94} 
If 2H ZnS crystals are cooled sufficiently fast to room temperature, they will retain
this 2H structure even though it is not the stable phase; presumably due to
the difficulty of shifting MLs to form the cubic (3C) structures. These
crystals are often further annealed for an hour at temperatures $500-800$ C,
thus perhaps providing the necessary activation energy to induce a solid-state
transformation from the 2H to a twinned 3C structure. This transformation has
been studied before, and many authors have concluded that deformation faulting
is the main mechanism responsible,~\cite{roth60,varn03b} at least for the case
of weak faulting.  

Engel and Needs~\cite{engel90b} performed a first-principles pseudopotential
energy calculation at $T=0$ to determine the coupling constants between MLs
up to separation three. They found an expression for the energy in the form 
\begin{eqnarray}
 {\mathcal H} = -\sum_{m=1}^{2} J_m \sum_{i=1}^{N}
	\Bigl[ \bigl( s_{i}-\frac{1}{2}\bigr) \bigl( s_{i+m}-\frac{1}{2}\bigr) \Bigr]~,
\label{eq:hexagonality}
\end{eqnarray}
where the $s_i \in\{0,1\}$ are \emph{inter-ML spins}~\cite{varn02} and $J_m$
the inter-ML coupling constants at separation $m$. They found $J_1 > 0$ and
$J_2 < 0$, but much smaller in magnitude than $J_1$. All other couplings
between MLs were found to be negligible. 

Our approach is to assume the following very simple model describes the
2H $\rightarrow$ 3C transformation in ZnS. 

\begin{trivlist}

\item (i) {\it The energetics of ML stacking in ZnS is describable by a
one-dimensional Ising chain with a nearest-neighbor interaction Hamiltonian
${\mathcal H}$ between MLs}. Further, we assume that entropic effects are small
at all temperatures of interest, so that the free energy $F = {\mathcal H}$.  

\item (ii) {\it The effective coupling $J_1(T)$ between adjacent MLs is
temperature dependent, being positive at temperatures below the transition,
$T_c \approx 1000$ C, and negative above.} While little is known about the 
temperature dependence of the effective couplings $J_i(T)$,~\cite{engel90a} 
this assumption gives the correct stable phases for ZnS (2H and 3C) above and below the
the transition temperature.  

\item (iii) {\it There is only one faulting mechanism---deformation
faulting---and it is driven by Glauber dynamics.} Deformation
faulting~\cite{varn03a} occurs in close-packed crystals when there is a slip
by a non-Bravais lattice vector between adjacent MLs. In terms of the spin
sequence, this corresponds to flipping a single spin~\cite{varn03a} and so
the faulting reduces to basic Glauber dynamics.~\cite{glauber63}  

\item (iv) {\it The transformation is slow, sluggish, and random.} By
\emph{slow}, we mean that the time for a slip between MLs to occur is much
shorter than the interval between slips. By \emph{sluggish}, we mean that a
slip will only occur if it is energetically favorable. And by \emph{random},
we mean that each ML is equally likely to be selected for possible faulting,
without any coordination between MLs.  
Since a deformation fault occurs when one portion of the crystal slips relative
to the other, it is reasonable to assume that this happens infrequently and
only when the slip results in a reduction of the energy.
While the {\em mechanism} of faulting assumed here is expressly random, we
will show nonetheless that this leads to a crystal {\em structure} that has a nonrandom
fault distribution.  

\end{trivlist}

Our model is most similar to that introduced by Engel.~\cite{engel90a} He
assumed two kinds of faults, fast and slow, depending on the configuration
of the five-spin neighborhood centered about the candidate spin to be flipped.
Restricting the range of interaction to nearest-neighbors, as we do, only
has the effect of suppressing the slow fault mechanism.  

We simulate the solid-state transformation in ZnS by assuming an initial
configuration of a pure 2H crystal with $N = 1048577$ MLs, which is quickly brought
into a temperature range ($500-800$ C) so that the 3C structure becomes the
stable phase, but there is sufficient thermal energy for MLs to slip.  
We randomly choose a spin in the configuration and apply the \emph{update rule}
listed in Table~\ref{update}. We visit the spins in a round-robin fashion,
so that each is updated once and only once, until all have been. It is easy
to see that the update rule permits a spin to flip only once, if at all, and
so we lose no generality in this model by visiting each spin only once.

\begin{table}
\begin{center}
\caption{
The deformation-faulting update rule for ZnS crystals suggested by the Hamiltonian
$\mathcal{H}$ with $J_1(T) > 0$ and $J_2(T) \approx 0$. Only those spin flips
that lower the energy of the spin configuration are allowed. This is equivalent to
applying elementary cellular automaton (ECA) rule 232 to the spin neighborhood
$\eta_i^t = ( s_{i-1}^t, s_i^t , s_{i+1}^t )$. $s_i^{t+1}$ gives the value assigned to
the center spin when it is selected for updating. The rule is applied
asynchronously to randomly selected sites.
}
\label{update}
\vspace{5 mm}
\begin{tabular}{l|cccccccc}
\hline
\hline
$\eta_i^t$ & $111$ & $110$ & $101$ & $100$ & $011$ & $010$ & $001$ & $000$ \\
\hline
$s_i^{t+1}$ & $1$  & $1$  & $1$  & $0$ &  $1$  & $0$  & $0$  & $0$ \\
\hline
\hline 
\end{tabular}
\end{center}
\end{table}

Let us define the \emph{faulting parameter} $f$ as the ratio of number $n_{app}$
of times the update rule has been applied to a crystal of size $N$:
$f = n_{app} / N$. The faulting parameter provides a useful
index of the amount of faulting the crystal has been subjected to or,
equivalently, of the stage of the transformation. That is, one can
determine the crystal structure at various stages along the transformation
by selecting different values for $f \in [0,1]$.
This algorithm is formally equivalent to applying elementary cellular automaton
rule $232$ asynchronously to a fraction $f$ of randomly chosen spins
(but no spin visited more than once). We then
call this model {\em asynchronous cellular automaton 232} (\ACA). Note that
the update rule is spin-inversion symmetric, in accord with the Hamiltonian.
Models of this basic flavor have been used to study systems other than
polytypism, such as cluster growth and phase separation in one
dimension,~\cite{privman92a,privman92b} diffusion-reaction problems in one
dimension,~\cite{spouge88,racz85} and the voter model.~\cite{bennaim96}
The transformation under \ACA\ of spatial configurations as a function of
$f$ is shown in Fig. \ref{fig:aca232sptmdiag},
starting from the period-two (2H) configuration.

\begin{figure}
\begin{center}
\resizebox{!}{8.0cm}{\includegraphics{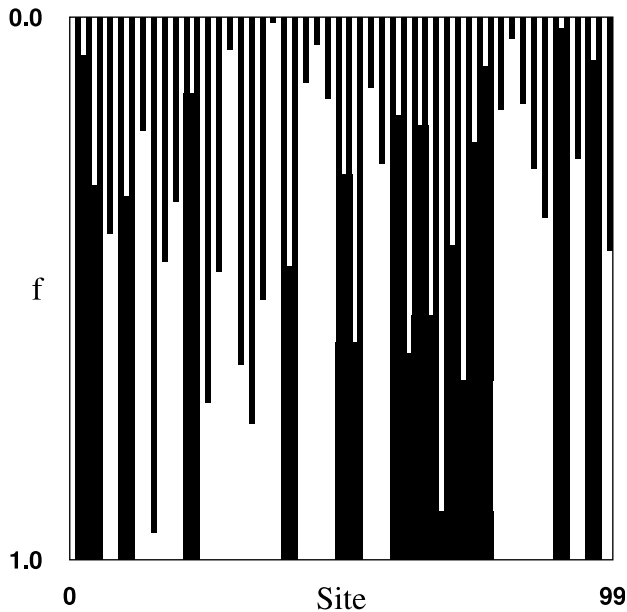}}
\end{center}
\caption{
  Configurations during the 2H $\rightarrow$ 3C transformation under
  asynchronous cellular automaton 232 on a $100$-site lattice with periodic
  boundary conditions. The vertical axis shows the degree $f$ of faulting and
  the horizontal axis is the spatial configuration, with black squares
  indicating $s_i=1$ inter-ML spins and white squares $s_i=0$ inter-ML spins.
  Initially, the crystal is in the 2H configuration as indicated by the
  alternating black and white squares at the top of the diagram ($f=0$).
  The faulting proceeds in a round-robin fashion until all sites are visited
  once. At the bottom of the diagram, ($f=1$), the crystal is fully faulted
  and there are only odd-spin domains (see text) with each domain having at
  least three spins.  
  }
\label{fig:aca232sptmdiag}
\end{figure}

\begin{figure}
\begin{center}
\resizebox{!}{12.0cm}{\includegraphics{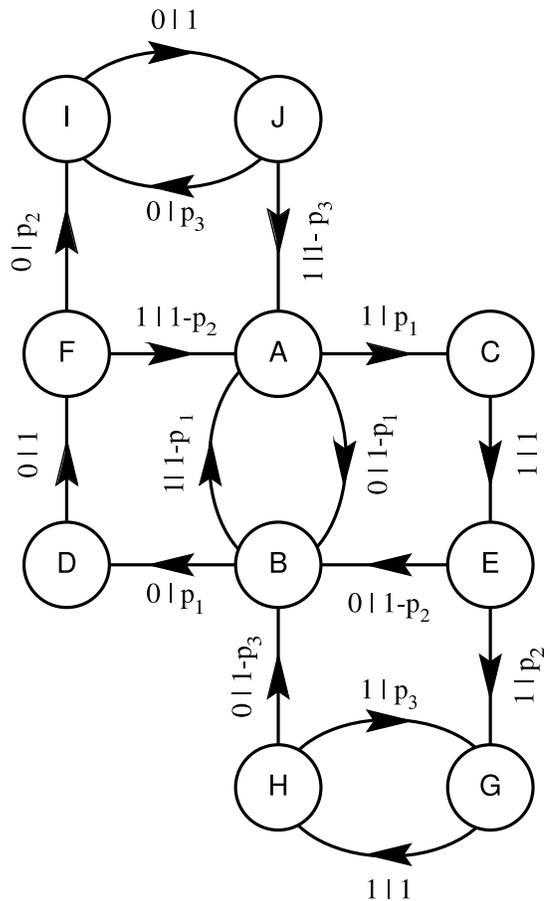}}
\end{center}
\caption{The reconstructed \eM\ that describes the architecture of the
  stacking process of a faulted 2H crystal across the range of possible fault parameters
  $f \in [0,1]$. The nodes represent causal states and the directed arcs,
  transitions between them. The edge labels $s|p$ indicate that a transition
  occurs from one causal state to another on symbol $s$ with probability $p$.  
  }
\label{fig:emachine.def.10}
\end{figure}

We then extract a structural description of a spin configuration in the form of
an \eM, as a function of $f$, using the causal-state splitting reconstruction
(CSSR)~\cite{Shal02a} algorithm.  
Specifically, we considered crystals with various amounts of faulting 
$f \in [0,1]$ in increments of $\Delta f = 0.10$
and performed \eM\ reconstruction for each $f$.  
We limit the number of causal states in the
reconstructed \eM\ by requiring that the addition of a new causal state reduce
the entropy rate by at least $1\%$.

\begin{figure}
\begin{center}
\resizebox{!}{5.2cm}{\includegraphics{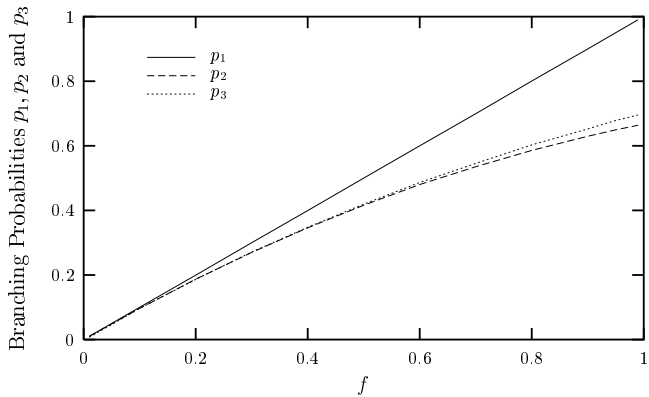}}
\end{center}
\caption[.]{The causal-state transition probabilities $p_1, p_2$, and $p_3$
  (see Fig. \ref{fig:emachine.def.10}) as a function of the faulting parameter
  $f$ for a crystal ($N = 1048577$ MLs) initially in a periodic 2H stacking
  sequence. The $p_i$ are estimated from reconstructed \eMs.
  }
\label{fig:fault.probs.2H}
\end{figure}

The \eM\ for a faulted 2H crystal appears in Fig.~\ref{fig:emachine.def.10}.
It has $10$ causal states and is a function of $3$ transition
probabilities---$p_1$, $p_2$, and $p_3$. These transition probabilities
are themselves a function of the faulting parameter $f$ and this dependence
is shown in Fig.~\ref{fig:fault.probs.2H}. 
For small values of the faulting parameter, $p_i \approx f$. This implies that
for small $f$, the \eM\ largely cycles between the causal states $\textsf{A}$
and $\textsf{B}$. This produces the spin sequence $\ldots 101010 \ldots$,
which we recognize as the 2H stacking sequence. So, for small faulting, the
original 2H crystal remains largely intact; as one expects. At the other
extreme, when $f=1$ (a fully faulted crystal) we have $p_1=1$, and there is
no longer a transition between the causal states $\textsf{A}$ and $\textsf{B}$.
Since there are no other \emph{causal-state cycles}~\cite{varn03a} that generate
the $\ldots 101010 \ldots$ spin sequence, we see that the original 2H structure
is eliminated. 

Since the \eM\ has spin-inversion symmetry---i.e., it is invariant
under $1 \Leftrightarrow 0$ exchange, we only need to examine half
of it to get an intuitive understanding of the structures it captures.
Let us consider only the causal states $\textsf{A}$, $\textsf{B}$,
$\textsf{D}$, $\textsf{F}$, $\textsf{I}$, and $\textsf{J}$. 
As stated before, the causal state cycle [$\textsf{A}\textsf{B}$]
generates the 2H structure. At $\textsf{B}$, however, it is possible to 
emit a $0$ and make a transition to $\textsf{D}$. This implies that
the spin history upon entering $\textsf{D}$ is 100 and, conversely,
having seen the spin sequence $100$ uniquely sets the \eM\ in
$\textsf{D}$. At this point, we see that $\textsf{D}$ must make a
transition to
$\textsf{F}$ on spin $0$. That is, the sequence $1001$ does not occur.
At $\textsf{F}$, the stacking process may either enter $\textsf{A}$
on a $1$ giving the cumulative spin sequence $10001$ or may proceed to
$\textsf{I}$ on a $0$. At $\textsf{I}$, the process must see another
$0$ and transition to $\textsf{J}$, giving the cumulative spin sequence
of $100000$. This implies that the spin sequence $100001$ likewise does
not occur. In fact, we see that the causal states $\textsf{D}$ and
$\textsf{I}$ force generating a $0$ as the next spin and, thus,
prevent sequences where two or four $0$s are sandwiched between two
$1$s. Similar reasoning shows that {\em any} even number of $0$s
($1$s) cannot be trapped between two $1$s ($0$s). In other words,
the sets of sequences $\lbrace 10^{2n}1 , n = 1, 2, 3, \ldots \rbrace$
and $\lbrace 01^{2n}0 , n = 1, 2, 3, \ldots \rbrace$, and any
configurations that contain them, are forbidden. Thus, starting from a
short-range crystalline structure (2H or $\ldots 010101 \ldots$),
annealing has produced complex spatial structures of infinite range.
 
We can understand how annealing accomplished this increase in structural
complexity in the following way. Define a {\em spin domain} as a
sequence of $k$ consecutive like spins sandwiched between two unlike
spins, where $k$ is a positive integer. In the original 2H structure,
there are no even-$k$ spin domains. The update rule has the effect of
joining two spin domains of the same spin by eliminating the $k=1$
spin domain separating them. Therefore, the resulting domain is of
size $k_{new} = k_l + k_r + 1$, where $k_l$ ($k_r$) is the $k$ value
of the spin domain to the left (right) of the flipped spin. The spin
that is flipped also contributes a site to the new spin domain. Thus,
if $k_l$ and $k_r$ are initially odd, $k_{new}$ is also. Since an even
spin domain cannot appear, as one scans the crystal one must remember
the number of consecutive like spins in order to determine the
admissibility of the next spin value. In general, this can require
remembering an indefinite number of previous spins. In this way, the process
has an infinite memory, so we say that the \emph{memory length} \Range\ is
infinite.\cite{varn03a} Put another way, spins arbitrarily far apart may
contain information about each other not contained by the intervening spins.
Symbolic dynamics calls such a process
{\em strictly sofic}.~\cite{Weis73} The most significant consequence of
being sofic is that no finite-order Markov process can describe the
stacking sequence. Notably, this suggests that previous efforts to
describe annealed, disordered crystals that assume some sort of Markov
model~\cite{jagodzinski49a,jagodzinski49b,roth60,pandey80a,pandey80b,pandey80c,varn02,varn03a,varn03b} 
are likely inadequate.  

As we noted earlier, the structure for a completely transformed crystal
occurs for $f=1$. Then the shortest 
causal-state cycle is [$\textsf{FACEBD}$] and generates
the spin configuration $\ldots 111000 \ldots$. There can be forays
into the causal states $\textsf{I}$ and $\textsf{J}$ or $\textsf{G}$
and $\textsf{H}$, but these only serve to increase the spin domains
by an even number. Thus, the final crystal is a
{\em disordered, twinned 3C crystal with only odd spin domains.}
Notably, some time ago, Mardix~\cite{mardix86} performed a statistical
analysis of the known 
crystalline polytypes in ZnS and found a significant bias toward odd number
Zhdanov elements in the configurations. These correspond to regions of
odd spin domains, in agreement with odd spin domains predicted by \ACA\
for disordered ZnS crystals.  

It is useful to inquire into the origin of the infinite memory length.
Clearly it does not derive directly from the interaction Hamiltonian,
which has an interaction range of one. In fact, it comes from the
repeated interaction of the stochastic (asynchronous) dynamics and the
structural constraints imposed by the allowed (deformation) faulting
mechanism. When a spin is flipped, information is lost. That is, when
scanning the spin configuration, the process must remember the last
$01$ or $10$ pair. The local update rule acts to eliminate pairs of
this form and so requires the process to remember an increasingly
longer ``history'' of spins, as annealing progresses. 

\begin{figure}
\begin{center}
\resizebox{!}{5.2cm}{\includegraphics{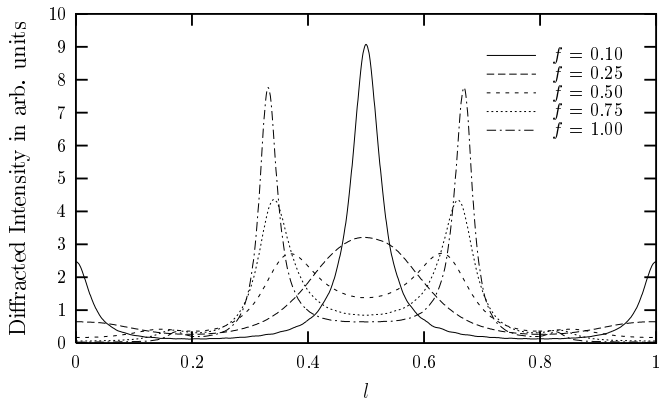}}
\end{center}
\caption[.]{The diffraction patterns along the $10.l$ row for partially
  annealed 2H crystals. All spectra, here and elsewhere (except for
  those from the fault model), are normalized to unity.
  }
\label{fig:diff.patterns}
\end{figure}

We now address the issue of nonrandom faulting. By examining the update rule
in Table~\ref{update}, we see that when a spin flips in the 
\ACA\ model, the two adjacent spins are necessarily prevented from also
flipping. This therefore precludes the possibility of two adjacent faults.  
This implies that faults are not random, since there must be at least a two 
ML separation between them. 
This is easily seen by examining a sequence of spins and the conditional
probabilities for faulting as implied by the \eM.  

Take a segment of an unfaulted 2H crystal as
$10101010$. Now introduce a fault (i.e. spin flip) in the spin sequence to get  
$10\underline{0}01010$, where the underline indicates the flipped spin. 
Now neither spin adjacent to the flipped spin can itself flip. However, the 
spins at a distance two or greater can. Let us consider additional 
spin flips at a separation of two, three, and four. We then have the
sequences $10\underline{0}0\underline{0}010$,
$10\underline{0}01\underline{1}10$, and
$10\underline{0}010\underline{0}0$ as possible stacking sequences, representing
deformation stacking faults at separations of two, three, and four.  
As before, the history $10\underline{0}$ requires the \eM\ to be in causal state    
$\textsf{D}$. The next spin must be 0, advancing the \eM\ to causal state 
$\textsf{F}$. Thus the causal state architecture of the \eM\ has 
prevent two adjacent faults, as required, and the spin sequence is now
$10\underline{0}0$. If the stacking sequence is to resume unfaulted, we expect
the next spin to be 1. From the \eM, however, the probability of 0, implying
a deformation fault separated from the first by two MLs, is $p_2$. 
That is, the conditional probability of a second fault occurring two MLs after the first
is $p_2$. 
Similar reasoning
shows the \eM\ specifies that the conditional probability of a deformation fault
occurring three and four MLs after the first is $p_1$. We find, therefore, that 
{\em the conditional probabilities of observing deformation faults at a
separation of one, two, and three  MLs are all different, even though
\ACA\ manifestly assumes only interactions between nearest neighbors.}
For small faulting, however, Fig.~\ref{fig:fault.probs.2H} shows that 
$p_1 \approx p_2$, so that this distinction is minor for faults at a separation of two and three. 

The above exercise demonstrates explicitly that the \eM\ can represent 
a nonrandom distribution of stacking faults 
and this distribution can arise from an interaction range of one combined with
a simple mechanism for annealing. \ACA\ then provides some theoretical
justification for models that assume structures with a nonrandom fault
distribution without appealing to any long-range interactions or 
coordination between faults.  

To see how well \ACA\ describes phase transitions in experimentally
observed in ZnS crystals, we calculated the diffraction spectrum along
the $10.l$ row for the faulted crystal for various values of the fault
parameter as shown in Fig.~\ref{fig:diff.patterns}.
(We employ the same procedure, definitions, and assumptions here as we
used previously.~\cite{varn03a}) 
Let $l$ be a dimensionless variable that indexes the magnitude of the
perpendicular component of the diffracted wave.
For small values of the faulting, the 2H Bragg peaks at $l=1/2$ and
$l = 1$ are widened considerably. As $f$ approaches $0.50$, the
enhancement in diffracted intensity at these positions has vanished
and there is instead significantly increased scattering near the
$l \approx 1/3$ and $l \approx 2/3$ positions, which is normally
associated with twinned 3C structure. There are also small rises in
the diffracted intensity at the $l \approx 1/6$ and $l \approx 5/6$
positions, which is often considered a sign of 6H structure. As $f$
approaches $1.0$, the peaks at $l \approx 1/3$ and $l \approx 2/3$
sharpen and the diffracted intensity at the $l \approx 1/6$ and
$l \approx 5/6$ positions weaken. For $f=1$, the crystal has found
a local minimum of the Hamiltonian, and no further transformation
occurs. The crystal is now a disordered, twinned 3C structure.

Notice also that it had been assumed previously that the small rises
in the intensity at $l \approx 1/6$ and $l \approx 5/6$ signaled that
parts of the crystal were faulting into the 6H structure, but we see
that no special assumption of this kind is needed. Rather, the spin
sequence $000111$ (i.e. the 6H structure) occurs quite naturally as
the 2H-crystal deformation faults, giving the enhanced diffracted
intensity at these positions.  

\begin{figure}
\begin{center}
\resizebox{!}{5.2cm}{\includegraphics{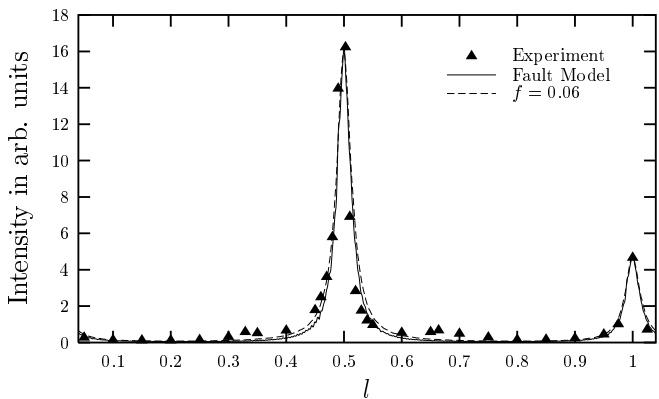}}
\end{center}
\caption[.]{A comparison of an experimental
  diffractogram~\cite{sebastian94,varn02,varn03b} along the $10.l$ row
  (triangles) with the
  fault model (solid line)~\cite{varn02} and \ACA\ (dotted line) at 
  $f=0.06$. Except for some minor differences, \ACA\ gives the same
  description as the fault model. We find the profile
  ${\cal R}$-factor~\cite{varn03b} between experiment and \ACA\ to be
  $29\%$. We see that \ACA\ describes the Bragg peaks at
  $l \approx 0.5$ and $l \approx 1$ well, but misses the small rise in
  intensity at $l \approx 0.67$. This likely results from 3C structure,
  so it is not surprising that it is absent the \ACA\ spectrum at
  low $f$.  
  }
\label{fig:SK134.1b.cor.ext.3.new}
\end{figure}

We now compare the predictions from our simple model with two experimental
ZnS diffraction spectra we have previously treated~\cite{varn02,varn03b}:
the first from an unannealed sample; the second, from an annealed
sample. Figure~\ref{fig:SK134.1b.cor.ext.3.new} shows the spectrum
from an as-grown, disordered 2H crystal. Our previous analysis showed
that, while there were displacement faults, growth faults, and some 3C
structure, the dominant faulting was due to deformation. It is
reasonable, then, to ask if \ACA\ can reproduce this spectrum. For
$f=0.06$ \ACA\ does reproduce the Bragg-like peaks at $l \approx 1/2$ and
$l \approx 1$ rather well. The small rise in diffuse scattering at
$l \approx 2/3$ is not recovered, nor should one expect it to be. This
feature comes from the small amount of 3C structure present and, as we
saw above, \ACA\ is not able to reproduce this. That it does not lends
further strength to our earlier interpretation~\cite{varn02} that a small
amount of 3C structure is present. In fact, we find that for small faulting,
\ACA\ gives nearly the same results as the fault model, as can
be seen in Fig.~\ref{fig:SK134.1b.cor.ext.3.new}.   

Figure~\ref{fig:SK135.1b.phtrans} compares an experimental ZnS
diffraction spectrum from a 2H crystal annealed at $500$ C for 1 h
with an $f = 85\%$ deformation-faulted 2H simulated diffraction spectra.
The agreement is excellent. The deformation-faulted crystal
reproduces the two  disordered 3C peaks in the experimental spectrum
at $l \approx -2/3$ and $l \approx -1/3$, as well as the diffuse
scattering between these diffraction maxima. Even the slight
enhancement in diffraction at $l \approx -1/6$ and $l \approx 1/6$
is captured (see inset). Our deformation-faulting model gives a
stacking configuration that is spin symmetric and so it is unable to
account for the relative difference in intensity between the two peaks
at $l \approx -2/3$ and $l \approx -1/3$. There is a question here
concerning the quality of the data,~\cite{varn03b} and so the difference
in intensities may be an experimental artifact. Note that the
diffracted intensity between the two Bragg-like peaks 
at $l \approx -2/3$ and $l \approx -1/3$ 
is especially well represented by
\ACA. This contrasts with the fault model which simply cannot recover
the diffuse scattering.   

\begin{figure}
\begin{center}
\resizebox{!}{5.2cm}{\includegraphics{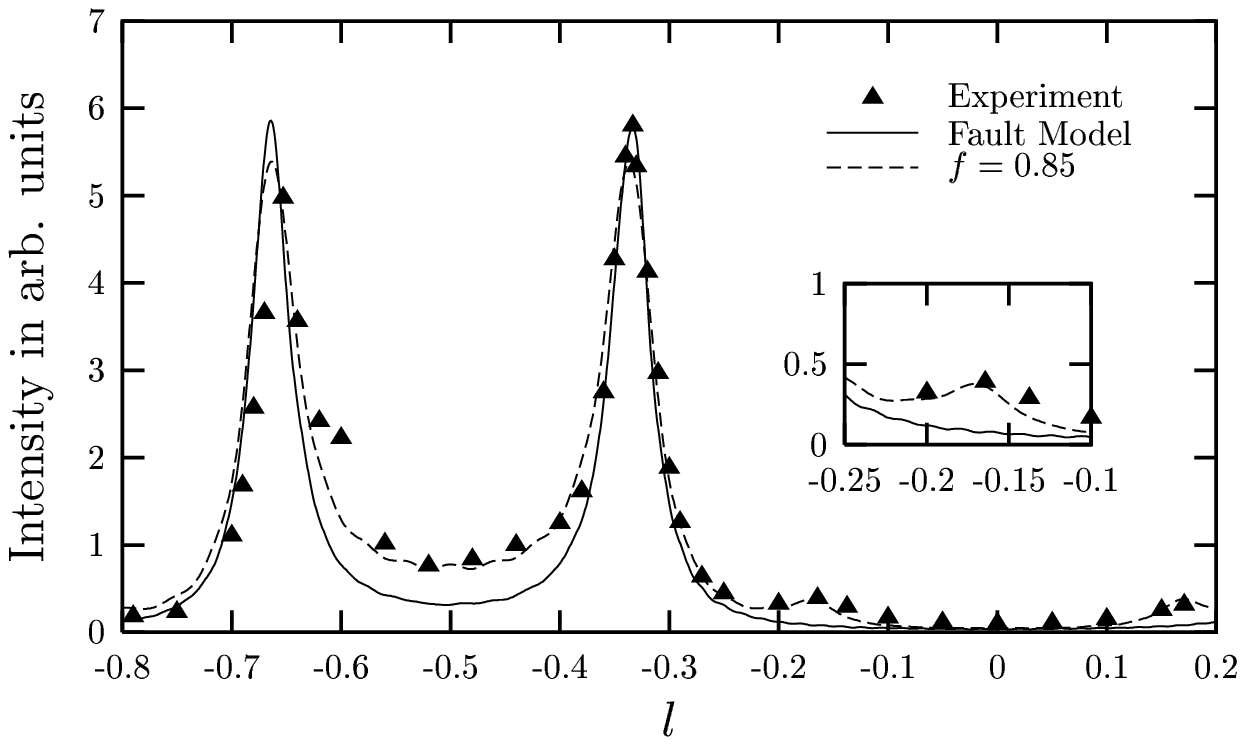}}
\end{center}
\caption[.]{A comparison of an experimental
  diffractogram~\cite{sebastian94,varn02} along the $10.l$ row
  (triangles) with the
  fault model (solid line)~\cite{sebastian94,varn02} and \ACA\ (dashed line) at 
  $f=0.85$. With a profile ${\cal R}$-factor of $18\%$
  (compare $13\%$ for \eMSR\ ~\cite{varn03b} and $33\%$ for the fault model),
  \ACA\ captures a surprisingly large amount of the stacking structure.
  Notice particularly that the inset shows that \ACA\ nicely reproduces
  the small rise in diffracted intensity at $l \approx -0.16$.  
  }
\label{fig:SK135.1b.phtrans}
\end{figure}

There are however, features seen in diffractograms of partially
annealed ZnS crystals that are not observed in the simulated
diffractograms of Fig.~\ref{fig:diff.patterns}. Namely, the
diffractograms of some partially transformed ZnS crystals, as in
Fig. 9 of Ref.~\citep{varn03b}, show enhancement simultaneously at
values of $l$ associated with the 2H structure, namely $l \approx 1/2$
and $l \approx 1$, as well as at $l$ values associated with twinned 3C
structure ($l \approx 1/3$ and $l \approx 2/3$). In our simulation
studies, the strong 2H reflections at $l \approx 1/2$ and $l \approx 1$
disappear before the 3C reflections at $l \approx 1/3$ and
$l \approx 2/3$ start becoming prominent. \ACA\ therefore has
difficulty even qualitatively explaining such spectra and one is
led to entertain models more complicated than \ACA. 

We introduced a very simple model, called \ACA\ due to its equivalence
with elementary cellular automaton rule 232, to study the 2H
$\rightarrow$ 3C phase transition in polytypic ZnS crystals. We assumed
only nearest-neighbor interactions between MLs and thus a local,
though asynchronously applied, faulting mechanism. We simulated a
solid-state transition during annealing of the crystal and found a
description of the stacking structure for all amounts of
faulting in the form of an \eM. 
The resulting \eM\ implied that, despite interactions only between nearest neighbors,
the structure of the crystal had a nonrandom distribution of faults.  
The \eM\ also shows an infinite
memory length, which means that arbitrarily separated spins share
nonredundant information. The corresponding symbolic dynamics
is strictly sofic and this implies that no finite-order Markov process
can completely specify the stacking in these faulted crystals.
Intuitively, this arises because only odd-length spin domains are
allowed, if the crystal starts from a perfect 2H specimen. 
The infinite memory length process discovered here contrasts with the $r=3$ memory length
we found by analyzing disordered ZnS crystals using \eMSR.~\cite{varn03b} This 
suggests that while the $r=3$ \eM\ captures much of the stacking structure
present, it is nonetheless only an approximation to a more structurally
complex stacking process. The relation between the \eM\ found in  
these simulation studies and that determined by \eMSR\ from experimental
spectra is currently a subject of investigation. Since even simple, 
annealed crystals probably have an infinite memory length, this
further suggests that the Markov approximations used in the \eMSR\ algorithm
should be circumvented and a more general spectral reconstruction technique is 
warranted. This, too, is a current subject of research.        
Finally, we compared
diffraction spectra calculated from \ACA\ to previously analyzed ZnS
spectra and found good quantitative agreement.

While not included here, our studies also considered different initial
stacking sequences. We found that often the memory length was infinite
after annealing and that the structure of the final faulted crystal was
highly dependent on the initial stacking structure. Although phase
transitions in actual ZnS crystals undoubtedly are more complicated, 
our simple model captures much of the noncritical organization they
undergo. Also, considering its simplicity, and the fact that similar
models have been used to study other systems, the results presented
here should have applicability far beyond the realm of polytypism. 

\begin{acknowledgments}

We thank Geoff Canright and Eric Smith for useful conversations. 
This work was supported at the Santa Fe Institute under the Networks
Dynamics Program funded by the Intel Corporation and under the
Computation, Dynamics, and Inference Program via SFI's core grants from
the National Science and MacArthur Foundations. Direct support was
provided by DARPA Agreement F30602-00-2-0583.    

\end{acknowledgments}

\bibliography{master.refs}

\begin{thebibliography}{39}
\expandafter\ifx\csname natexlab\endcsname\relax\def\natexlab#1{#1}\fi
\expandafter\ifx\csname bibnamefont\endcsname\relax
  \def\bibnamefont#1{#1}\fi
\expandafter\ifx\csname bibfnamefont\endcsname\relax
  \def\bibfnamefont#1{#1}\fi
\expandafter\ifx\csname citenamefont\endcsname\relax
  \def\citenamefont#1{#1}\fi
\expandafter\ifx\csname url\endcsname\relax
  \def\url#1{\texttt{#1}}\fi
\expandafter\ifx\csname urlprefix\endcsname\relax\def\urlprefix{URL }\fi
\providecommand{\bibinfo}[2]{#2}
\providecommand{\eprint}[2][]{\url{#2}}

\bibitem[{\citenamefont{Hendricks and Teller}(1942)}]{hendricks42}
\bibinfo{author}{\bibfnamefont{S.}~\bibnamefont{Hendricks}} \bibnamefont{and}
  \bibinfo{author}{\bibfnamefont{E.}~\bibnamefont{Teller}},
  \bibinfo{journal}{J.\ Chem.\ Phys.} \textbf{\bibinfo{volume}{10}},
  \bibinfo{pages}{147} (\bibinfo{year}{1942}).

\bibitem[{\citenamefont{Jagodzinski}(1949{\natexlab{a}})}]{jagodzinski49a}
\bibinfo{author}{\bibfnamefont{H.}~\bibnamefont{Jagodzinski}},
  \bibinfo{journal}{Acta Crystallogr.} \textbf{\bibinfo{volume}{2}},
  \bibinfo{pages}{201} (\bibinfo{year}{1949}{\natexlab{a}}).

\bibitem[{\citenamefont{Pandey et~al.}(1980{\natexlab{a}})\citenamefont{Pandey,
  Lele, and Krishna}}]{pandey80a}
\bibinfo{author}{\bibfnamefont{D.}~\bibnamefont{Pandey}},
  \bibinfo{author}{\bibfnamefont{S.}~\bibnamefont{Lele}}, \bibnamefont{and}
  \bibinfo{author}{\bibfnamefont{P.}~\bibnamefont{Krishna}},
  \bibinfo{journal}{Proc.\ R.\ Soc.\ London Ser.\ A}
  \textbf{\bibinfo{volume}{369}}, \bibinfo{pages}{435}
  (\bibinfo{year}{1980}{\natexlab{a}}).

\bibitem[{\citenamefont{Pandey et~al.}(1980{\natexlab{b}})\citenamefont{Pandey,
  Lele, and Krishna}}]{pandey80b}
\bibinfo{author}{\bibfnamefont{D.}~\bibnamefont{Pandey}},
  \bibinfo{author}{\bibfnamefont{S.}~\bibnamefont{Lele}}, \bibnamefont{and}
  \bibinfo{author}{\bibfnamefont{P.}~\bibnamefont{Krishna}},
  \bibinfo{journal}{Proc.\ R.\ Soc.\ London Ser.\ A}
  \textbf{\bibinfo{volume}{369}}, \bibinfo{pages}{451}
  (\bibinfo{year}{1980}{\natexlab{b}}).

\bibitem[{\citenamefont{Pandey et~al.}(1980{\natexlab{c}})\citenamefont{Pandey,
  Lele, and Krishna}}]{pandey80c}
\bibinfo{author}{\bibfnamefont{D.}~\bibnamefont{Pandey}},
  \bibinfo{author}{\bibfnamefont{S.}~\bibnamefont{Lele}}, \bibnamefont{and}
  \bibinfo{author}{\bibfnamefont{P.}~\bibnamefont{Krishna}},
  \bibinfo{journal}{Proc.\ R.\ Soc.\ London Ser.\ A}
  \textbf{\bibinfo{volume}{369}}, \bibinfo{pages}{463}
  (\bibinfo{year}{1980}{\natexlab{c}}).

\bibitem[{\citenamefont{Berliner and Werner}(1986)}]{berliner86}
\bibinfo{author}{\bibfnamefont{R.}~\bibnamefont{Berliner}} \bibnamefont{and}
  \bibinfo{author}{\bibfnamefont{S.}~\bibnamefont{Werner}},
  \bibinfo{journal}{Phys.\ Rev.\ B} \textbf{\bibinfo{volume}{34}},
  \bibinfo{pages}{3586} (\bibinfo{year}{1986}).

\bibitem[{\citenamefont{Gosk}(2001)}]{gosk01}
\bibinfo{author}{\bibfnamefont{J.~B.} \bibnamefont{Gosk}},
  \bibinfo{journal}{Crys.\ Res.\ Tech.} \textbf{\bibinfo{volume}{36}},
  \bibinfo{pages}{197} (\bibinfo{year}{2001}).

\bibitem[{\citenamefont{Varn and Canright}(2001)}]{varn01a}
\bibinfo{author}{\bibfnamefont{D.~P.} \bibnamefont{Varn}} \bibnamefont{and}
  \bibinfo{author}{\bibfnamefont{G.~S.} \bibnamefont{Canright}},
  \bibinfo{journal}{Acta Crystallogr.\ Sec.\ A} \textbf{\bibinfo{volume}{57}},
  \bibinfo{pages}{4} (\bibinfo{year}{2001}).

\bibitem[{\citenamefont{Sebastian and Krishna}(1994)}]{sebastian94}
\bibinfo{author}{\bibfnamefont{M.~T.} \bibnamefont{Sebastian}}
  \bibnamefont{and} \bibinfo{author}{\bibfnamefont{P.}~\bibnamefont{Krishna}},
  \emph{\bibinfo{title}{Random, Non-Random and Periodic Faulting in Crystals}}
  (\bibinfo{publisher}{Gordon and Breach}, \bibinfo{year}{1994}).

\bibitem[{\citenamefont{Ellis et~al.}(1958)\citenamefont{Ellis, Herman,
  Loebner, Merz, Struck, and White}}]{ellis58}
\bibinfo{author}{\bibfnamefont{S.~G.} \bibnamefont{Ellis}},
  \bibinfo{author}{\bibfnamefont{F.}~\bibnamefont{Herman}},
  \bibinfo{author}{\bibfnamefont{E.~E.} \bibnamefont{Loebner}},
  \bibinfo{author}{\bibfnamefont{W.~J.} \bibnamefont{Merz}},
  \bibinfo{author}{\bibfnamefont{C.~W.} \bibnamefont{Struck}},
  \bibnamefont{and} \bibinfo{author}{\bibfnamefont{J.~G.} \bibnamefont{White}},
  \bibinfo{journal}{Phys.\ Rev.} \textbf{\bibinfo{volume}{109}},
  \bibinfo{pages}{1860} (\bibinfo{year}{1958}).

\bibitem[{\citenamefont{Varn et~al.}(2002)\citenamefont{Varn, Canright, and
  Crutchfield}}]{varn02}
\bibinfo{author}{\bibfnamefont{D.~P.} \bibnamefont{Varn}},
  \bibinfo{author}{\bibfnamefont{G.~S.} \bibnamefont{Canright}},
  \bibnamefont{and} \bibinfo{author}{\bibfnamefont{J.~P.}
  \bibnamefont{Crutchfield}}, \bibinfo{journal}{Phys.\ Rev.\ B.}
  \textbf{\bibinfo{volume}{66}}, \bibinfo{pages}{156} (\bibinfo{year}{2002}).

\bibitem[{\citenamefont{Varn et~al.}(2003{\natexlab{a}})\citenamefont{Varn,
  Canright, and Crutchfield}}]{varn03a}
\bibinfo{author}{\bibfnamefont{D.~P.} \bibnamefont{Varn}},
  \bibinfo{author}{\bibfnamefont{G.~S.} \bibnamefont{Canright}},
  \bibnamefont{and} \bibinfo{author}{\bibfnamefont{J.~P.}
  \bibnamefont{Crutchfield}}, \bibinfo{journal}{Phys.\ Rev.\ B} p.
  \bibinfo{pages}{submitted} (\bibinfo{year}{2003}{\natexlab{a}}).

\bibitem[{\citenamefont{Varn et~al.}(2003{\natexlab{b}})\citenamefont{Varn,
  Canright, and Crutchfield}}]{varn03b}
\bibinfo{author}{\bibfnamefont{D.~P.} \bibnamefont{Varn}},
  \bibinfo{author}{\bibfnamefont{G.~S.} \bibnamefont{Canright}},
  \bibnamefont{and} \bibinfo{author}{\bibfnamefont{J.~P.}
  \bibnamefont{Crutchfield}}, \bibinfo{journal}{Phys.\ Rev.\ B} p.
  \bibinfo{pages}{submitted} (\bibinfo{year}{2003}{\natexlab{b}}).

\bibitem[{\citenamefont{Trigunayat}(1991)}]{trigunyat91}
\bibinfo{author}{\bibfnamefont{G.~C.} \bibnamefont{Trigunayat}},
  \bibinfo{journal}{Solid State Ionics} \textbf{\bibinfo{volume}{48}},
  \bibinfo{pages}{3} (\bibinfo{year}{1991}).

\bibitem[{\citenamefont{Engel and Needs}(1990)}]{engel90b}
\bibinfo{author}{\bibfnamefont{G.~E.} \bibnamefont{Engel}} \bibnamefont{and}
  \bibinfo{author}{\bibfnamefont{R.~J.} \bibnamefont{Needs}},
  \bibinfo{journal}{J.\ Phys.\ Cond.\ Mat.} \textbf{\bibinfo{volume}{2}},
  \bibinfo{pages}{367} (\bibinfo{year}{1990}).

\bibitem[{\citenamefont{Cheng et~al.}(1987)\citenamefont{Cheng, Needs, Heine,
  and Churcher}}]{cheng87}
\bibinfo{author}{\bibfnamefont{C.}~\bibnamefont{Cheng}},
  \bibinfo{author}{\bibfnamefont{R.~J.} \bibnamefont{Needs}},
  \bibinfo{author}{\bibfnamefont{V.}~\bibnamefont{Heine}}, \bibnamefont{and}
  \bibinfo{author}{\bibfnamefont{N.}~\bibnamefont{Churcher}},
  \bibinfo{journal}{Europhys.\ Lett.} \textbf{\bibinfo{volume}{4}},
  \bibinfo{pages}{475} (\bibinfo{year}{1987}).

\bibitem[{\citenamefont{Price and Yeomens}(1984)}]{price84}
\bibinfo{author}{\bibfnamefont{G.}~\bibnamefont{Price}} \bibnamefont{and}
  \bibinfo{author}{\bibfnamefont{J.}~\bibnamefont{Yeomens}},
  \bibinfo{journal}{Acta Crystallogr., Sec. B} \textbf{\bibinfo{volume}{40}},
  \bibinfo{pages}{448} (\bibinfo{year}{1984}).

\bibitem[{\citenamefont{Yeomans}(1988)}]{yeomens88}
\bibinfo{author}{\bibfnamefont{J.}~\bibnamefont{Yeomans}},
  \bibinfo{journal}{Solid State Physics} \textbf{\bibinfo{volume}{41}},
  \bibinfo{pages}{151} (\bibinfo{year}{1988}).

\bibitem[{\citenamefont{Frank}(1951)}]{frank51b}
\bibinfo{author}{\bibfnamefont{F.~C.} \bibnamefont{Frank}},
  \bibinfo{journal}{Philos. Mag.} \textbf{\bibinfo{volume}{42}},
  \bibinfo{pages}{1014} (\bibinfo{year}{1951}).

\bibitem[{\citenamefont{Crutchfield and Young}(1989)}]{crutchfield89}
\bibinfo{author}{\bibfnamefont{J.~P.} \bibnamefont{Crutchfield}}
  \bibnamefont{and} \bibinfo{author}{\bibfnamefont{K.}~\bibnamefont{Young}},
  \bibinfo{journal}{Phys.\ Rev.\ Lett.} \textbf{\bibinfo{volume}{63}},
  \bibinfo{pages}{105} (\bibinfo{year}{1989}).

\bibitem[{\citenamefont{Crutchfield and Shalizi}(1999)}]{Crut98d}
\bibinfo{author}{\bibfnamefont{J.~P.} \bibnamefont{Crutchfield}}
  \bibnamefont{and} \bibinfo{author}{\bibfnamefont{C.~R.}
  \bibnamefont{Shalizi}}, \bibinfo{journal}{Physical Review E}
  \textbf{\bibinfo{volume}{59}}, \bibinfo{pages}{275} (\bibinfo{year}{1999}).

\bibitem[{\citenamefont{Kabra and Pandey}(1988)}]{kabra88}
\bibinfo{author}{\bibfnamefont{V.~K.} \bibnamefont{Kabra}} \bibnamefont{and}
  \bibinfo{author}{\bibfnamefont{D.}~\bibnamefont{Pandey}},
  \bibinfo{journal}{Phys.\ Rev.\ Lett.} \textbf{\bibinfo{volume}{61}},
  \bibinfo{pages}{1493} (\bibinfo{year}{1988}).

\bibitem[{not()}]{note1}
\bibinfo{note}{We use the Ramsell notation to specify structures in
  close-packed crystals. 2H corresponds to hexagonal close-packed and 3C to
  cubic close-packed.}

\bibitem[{\citenamefont{Engel}(1990)}]{engel90a}
\bibinfo{author}{\bibfnamefont{G.~E.} \bibnamefont{Engel}},
  \bibinfo{journal}{J.\ Phys.\ Cond.\ Mat.} \textbf{\bibinfo{volume}{2}},
  \bibinfo{pages}{6905} (\bibinfo{year}{1990}).

\bibitem[{\citenamefont{Shrestha et~al.}(1996)\citenamefont{Shrestha, Tripathi,
  Kabra, and Pandey}}]{shrestha96a}
\bibinfo{author}{\bibfnamefont{S.~P.} \bibnamefont{Shrestha}},
  \bibinfo{author}{\bibfnamefont{V.}~\bibnamefont{Tripathi}},
  \bibinfo{author}{\bibfnamefont{V.~K.} \bibnamefont{Kabra}}, \bibnamefont{and}
  \bibinfo{author}{\bibfnamefont{D.}~\bibnamefont{Pandey}},
  \bibinfo{journal}{Acta Mater.} \textbf{\bibinfo{volume}{44}},
  \bibinfo{pages}{4937} (\bibinfo{year}{1996}).

\bibitem[{\citenamefont{Sebastian and
  Krishna}(1987{\natexlab{a}})}]{sebastian87a}
\bibinfo{author}{\bibfnamefont{M.~T.} \bibnamefont{Sebastian}}
  \bibnamefont{and} \bibinfo{author}{\bibfnamefont{P.}~\bibnamefont{Krishna}},
  \bibinfo{journal}{Crys.\ Res.\ Tech.} \textbf{\bibinfo{volume}{22}},
  \bibinfo{pages}{929} (\bibinfo{year}{1987}{\natexlab{a}}).

\bibitem[{\citenamefont{Sebastian and
  Krishna}(1987{\natexlab{b}})}]{sebastian87b}
\bibinfo{author}{\bibfnamefont{M.~T.} \bibnamefont{Sebastian}}
  \bibnamefont{and} \bibinfo{author}{\bibfnamefont{P.}~\bibnamefont{Krishna}},
  \bibinfo{journal}{Crys.\ Res.\ Tech} \textbf{\bibinfo{volume}{22}},
  \bibinfo{pages}{1063} (\bibinfo{year}{1987}{\natexlab{b}}).

\bibitem[{\citenamefont{Gosk}(2000)}]{gosk00}
\bibinfo{author}{\bibfnamefont{J.~B.} \bibnamefont{Gosk}},
  \bibinfo{journal}{Crys.\ Res.\ Tech.} \textbf{\bibinfo{volume}{35}},
  \bibinfo{pages}{101} (\bibinfo{year}{2000}).

\bibitem[{\citenamefont{Roth}(1960)}]{roth60}
\bibinfo{author}{\bibfnamefont{W.~L.} \bibnamefont{Roth}}, \bibinfo{type}{Tech.
  Rep.} \bibinfo{number}{60-RL-2563M}, \bibinfo{institution}{General Electric
  Research}, \bibinfo{address}{Schenectady, New York} (\bibinfo{year}{1960}).

\bibitem[{\citenamefont{Glauber}(1963)}]{glauber63}
\bibinfo{author}{\bibfnamefont{R.~J.} \bibnamefont{Glauber}},
  \bibinfo{journal}{J. Math. Phys.} \textbf{\bibinfo{volume}{4}},
  \bibinfo{pages}{294} (\bibinfo{year}{1963}).

\bibitem[{\citenamefont{Privman}(1992{\natexlab{a}})}]{privman92a}
\bibinfo{author}{\bibfnamefont{V.}~\bibnamefont{Privman}}, \bibinfo{journal}{J.
  Stat. Phys.} \textbf{\bibinfo{volume}{69}}, \bibinfo{pages}{629}
  (\bibinfo{year}{1992}{\natexlab{a}}).

\bibitem[{\citenamefont{Privman}(1992{\natexlab{b}})}]{privman92b}
\bibinfo{author}{\bibfnamefont{V.}~\bibnamefont{Privman}},
  \bibinfo{journal}{Phys. Rev. Lett.} \textbf{\bibinfo{volume}{69}},
  \bibinfo{pages}{3686} (\bibinfo{year}{1992}{\natexlab{b}}).

\bibitem[{\citenamefont{Spouge}(1988)}]{spouge88}
\bibinfo{author}{\bibfnamefont{J.~L.} \bibnamefont{Spouge}},
  \bibinfo{journal}{Phys. Rev. Lett.} \textbf{\bibinfo{volume}{60}},
  \bibinfo{pages}{871} (\bibinfo{year}{1988}).

\bibitem[{\citenamefont{R\'{a}cz}(1985)}]{racz85}
\bibinfo{author}{\bibfnamefont{Z.}~\bibnamefont{R\'{a}cz}},
  \bibinfo{journal}{Phys. Rev. Lett.} \textbf{\bibinfo{volume}{55}},
  \bibinfo{pages}{1707} (\bibinfo{year}{1985}).

\bibitem[{\citenamefont{Ben-Naim et~al.}(1996)\citenamefont{Ben-Naim,
  Frachebourg, and Krapivsky}}]{bennaim96}
\bibinfo{author}{\bibfnamefont{E.}~\bibnamefont{Ben-Naim}},
  \bibinfo{author}{\bibfnamefont{L.}~\bibnamefont{Frachebourg}},
  \bibnamefont{and}
  \bibinfo{author}{\bibfnamefont{P.}~\bibnamefont{Krapivsky}},
  \bibinfo{journal}{Phys.\ Rev.\ E} \textbf{\bibinfo{volume}{53}},
  \bibinfo{pages}{3078} (\bibinfo{year}{1996}).

\bibitem[{\citenamefont{Shalizi et~al.}(2002)\citenamefont{Shalizi, Shalizi,
  and hfield}}]{Shal02a}
\bibinfo{author}{\bibfnamefont{C.~R.} \bibnamefont{Shalizi}},
  \bibinfo{author}{\bibfnamefont{K.~L.} \bibnamefont{Shalizi}},
  \bibnamefont{and} \bibinfo{author}{\bibfnamefont{J.~P.~C.}
  \bibnamefont{hfield}}, \bibinfo{journal}{Journal of Machine Learning
  Research} \textbf{\bibinfo{volume}{submitted}} (\bibinfo{year}{2002}),
  \bibinfo{note}{santa Fe Institute Working Paper 02-10-060;
  arXiv.org/abs/cs.LG/0210 025}.

\bibitem[{\citenamefont{Weiss}(1973)}]{Weis73}
\bibinfo{author}{\bibfnamefont{B.}~\bibnamefont{Weiss}},
  \bibinfo{journal}{Monastsh. Math.} \textbf{\bibinfo{volume}{77}},
  \bibinfo{pages}{462} (\bibinfo{year}{1973}).

\bibitem[{\citenamefont{Jagodzinski}(1949{\natexlab{b}})}]{jagodzinski49b}
\bibinfo{author}{\bibfnamefont{H.}~\bibnamefont{Jagodzinski}},
  \bibinfo{journal}{Acta Crystallogr.} \textbf{\bibinfo{volume}{2}},
  \bibinfo{pages}{208} (\bibinfo{year}{1949}{\natexlab{b}}).

\bibitem[{\citenamefont{Mardix}(1986)}]{mardix86}
\bibinfo{author}{\bibfnamefont{S.}~\bibnamefont{Mardix}},
  \bibinfo{journal}{Phys. Rev. B} \textbf{\bibinfo{volume}{33}},
  \bibinfo{pages}{8677} (\bibinfo{year}{1986}).

\end{thebibliography}

\end{document}